\documentclass{elsart}
\journal{New Astronomy}

\usepackage{natbib}

\usepackage{graphicx}
\usepackage{epsfig}

\usepackage{amssymb}

\begin{document}

\begin{frontmatter}


\title{Quasars as Probes of Late Reionization and Early Structure Formation}

\author{S.G. Djorgovski, M. Bogosavljevic, and A. Mahabal}
\address{Astronomy, MS 105-24, Caltech, Pasadena CA 91125\\
E-mail: [george,milan,aam]@astro.caltech.edu}

\begin{abstract}
Observations of QSOs at $z \sim 5.7 - 6.4$ show the appearance of
Gunn-Peterson troughs around $z \sim 6$, and a change in the slope of
the IGM optical depth $\tau(z)$ near $z \sim 5.5$.  These results are
interpreted as a signature of the end of the reionization era, which 
probably started at considerably higher redshifts.  However, there also
appears to be a substantial cosmic variance in the transmission of the
IGM, both along some lines of sight, and among different lines of sight,
in this intriguing redshift regime.  We suggest that this is indicative
of a spatially uneven reionization, possibly caused by the bias-driven
primordial clustering of the reionization sources.  There is also some
independent evidence for a strong clustering of QSOs at $z \sim 4 - 5$
and galaxies around them, supporting the idea of the strong biasing of
the first luminous sources at these redshifts.  Larger samples of
high-$z$ QSOs are needed in order to provide improved, statistically
significant constraints for the models of these phenomena.  We expect
that the Palomar-Quest (PQ) survey will soon provide a new set of QSOs
to be used as cosmological probes in this redshift regime.
\end{abstract}

\begin{keyword}
reionization \sep structure formation \sep quasars \sep cosmic variance

\end{keyword}

\end{frontmatter}

\section{Introduction}

Understanding of the cosmic reionization era -- from the appearance of the
first luminous sources perhaps at $z \sim 20$, to the end of the IGM phase
transition from neutral to essentially fully ionized hydrogen at $z \sim 6$
-- is now perhaps the focal arena of cosmology.  This field connects a
number of fundamental astrophysical processes, formation of the first stars,
galaxies, AGN, and large-scale structure, and their effect on the early IGM,
more than any other redshift range studied so far.  For recent
reviews see, e.g., \cite{loe+01, bar+01, mir+03, djo+05}.

There are so far three kinds of observational constraints on the nature and
extent of the reionization era.  Historically the first was the evidence of
the approach to the reionization, or its final stages, at $z \sim 6$ from
the spectra of QSOs in this redshift range
\cite{bec+01, fan+01, fan+03, djo+01, whi+03}.

The primary evidence is the existence of Gunn-Peterson (hereafter GP) 
\cite{gun+65} absorption troughs in the spectra of $z > 6$ QSOs.
Such features are now seen along all lines of sight at $z > 6$ where
adequate data exist.
There is also a growing evidence for a change in the slope of the optical
depth vs. redshift, $\tau(z)$~\cite{fan+02,cen+02,whi+03}
around $z \sim 5.5 - 6$, indicating that some qualitative change in the
physics and geometry of IGM occurs at these redshifts.
Transition from the thickening of the normal Ly$\alpha$ forest out to
$z \sim 5$ to the UV-opaque IGM at $z \sim 6$ is fairly dramatic (see Fig. 1).
As discoveries of more such objects continue \cite{fan+04},
spectroscopy of high-$z$ QSOs remains one of the principal empirical 
approaches to mapping of the final stages of reionization.

\begin{figure}[!t]
\centerline{\psfig{file=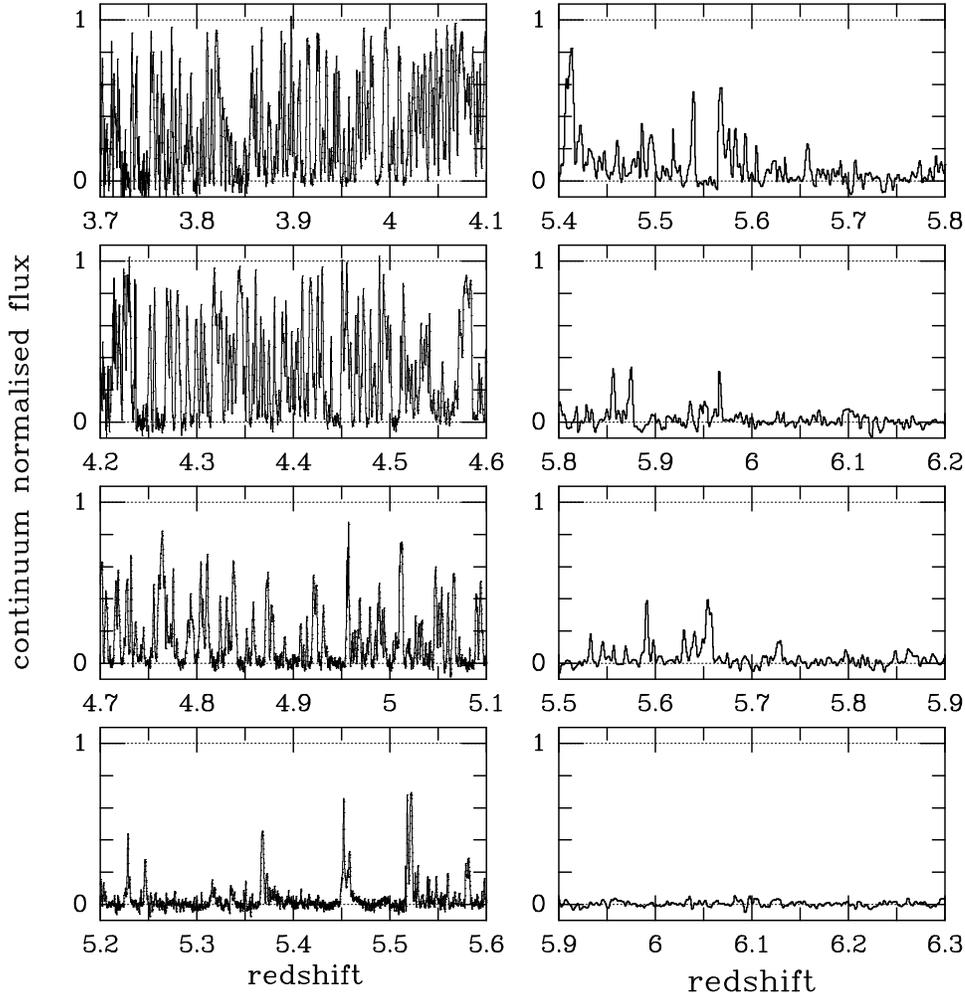,width=5.0in}}
\caption{
From the forest to the fog: continuum-normalized QSO absorption spectra at
progressively higher redshift windows.  The forest thickens dramatically at 
$z \sim 5.5$ and GP-type troughs are ubiquitous at $z \ge 6$.
Spectra of several high-$z$
SDSS QSOs, obtained at the Keck-II telescope with the ESI instrument were used.
From \cite{djo+02,djo+05}.
}
\label{}
\end{figure}

The picture was complicated by the evidence for an $early$ reionization at
$z \sim 10 - 20$ from the CMBR measurements by the WMAP satellite~\cite{kog+03}.
While the confirmation of this important result is still to come, we note
that it is not necessarily contradictory to the QSO results: the history of
the luminous sources and their effect on the IGM was probably highly complex,
and there was a finite time interval from the appearance of the first sources
of UV photons at the end of the cosmic dark ages, and the completion of the
reionization phase transition a few hundred million years later.

In parallel, a number of groups have studied star-forming galaxies at
$z \sim 6 - 7$, and measurements of the Ly$\alpha$ emission line luminosity
function evolution provide another useful observational constraint
\cite{mal+04,ste+05}.
While the QSO absorption
spectra probe the neutral hydrogen fraction regime $x_{H I} \leq 10^{-2}$,
this method is sensitive to the range $x_{H I} \sim 10^{-1} - 10^{0}$.

In the future, using H I hyperfine transition line at 21 cm promises to
be another powerful probe of the earlier stages of reionization, as
reviewed in a number of talks at this conference.  Detection and spectroscopy
of GRB afterglows from the first generations of massive stars will be another
spectacular method to study the astrophysics of reionization.

If, as both the data and theory now suggest, the distribution of the first
luminous sources was highly clustered, leading to a spatially uneven
reionization phase transition, direct measurements of the primordial 
clustering would provide a complementary insight into the overall picture.
Luminous QSOs at these redshifts are potentially a powerful probe of such
bias-driven clustering; see, e.g.,~\cite{djo+05} and references therein.

Here we review some of the current evidence for a spatially inhomogeneous
reionization, and the possible uses of high-$z$ QSOs to study the late
stages of this fundamental cosmological era.  In some ways, QSOs provide
unique observational constraints, and are in many ways complementary to
other approaches.  We also outline some possibilities for the future work.

\section{Evidence for a Clumpy Reionization?}

The reionization era must be extended in redshift, from the time the first
luminous sources turn on, to the final percolation of the reionized bubbles
universe-wide.  At any given intermediate redshift, there will be some
spatially uneven distribution of the neutral and ionized IGM, whose
characteristic physical scales and topology will depend on the clustering
and luminosity histories of the local reionization sources.  In principle,
such variations in the geometry/topology of reionization should be
detectable as variations in the IGM transmission, both along and among
different lines of sight to sources still within the incomplete reionization
regime.

Such variations are indeed observed, and can in principle be used to place
quantitative constraints on theoretical models~\cite{djo+02,djo+05,bog+06};
see Figs. 2 and 3.  Measurements of the Ly$\alpha$ and Ly$\beta$ optical
depth to a number of SDSS high-$z$ QSOs suggest a comparable cosmic
variance in the IGM transmission in this redshift regime \cite{str+05}.

\begin{figure}[!t]
\centerline{\psfig{file=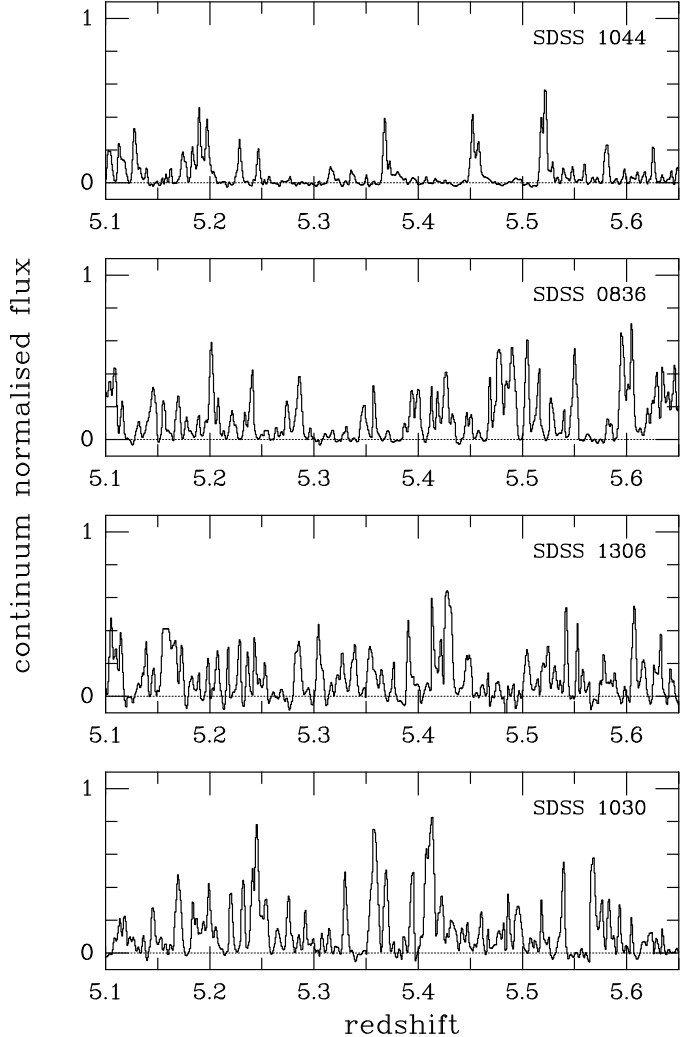,width=3.5in}}
\caption{
A substantial variety in the IGM absorption along 4 different QSO lines of
sight is apparent in these continuum-normalized spectra covering the same
redshift window, which just follows the reionization epoch.  This variance
may be a residual effect of an uneven end to reionization,
as expected from the biased structure formation picture.
From \cite{djo+02,djo+05}.
}
\label{}
\end{figure}

What is interesting here is that there may be an apparent increase in the
cosmic variance of the IGM transmission as one approaches the ostensible 
end-of-the-reionization redshift, $z \sim 6$.  The variance is impossible
to measure as the flux disappears at higher redshifts, but of course a
substantial variation in the neutral hydrogen fraction could be present
in the optically thick regime; this calls for some alternative measurement
methodology.

\begin{figure}[!t]
\centerline{\psfig{file=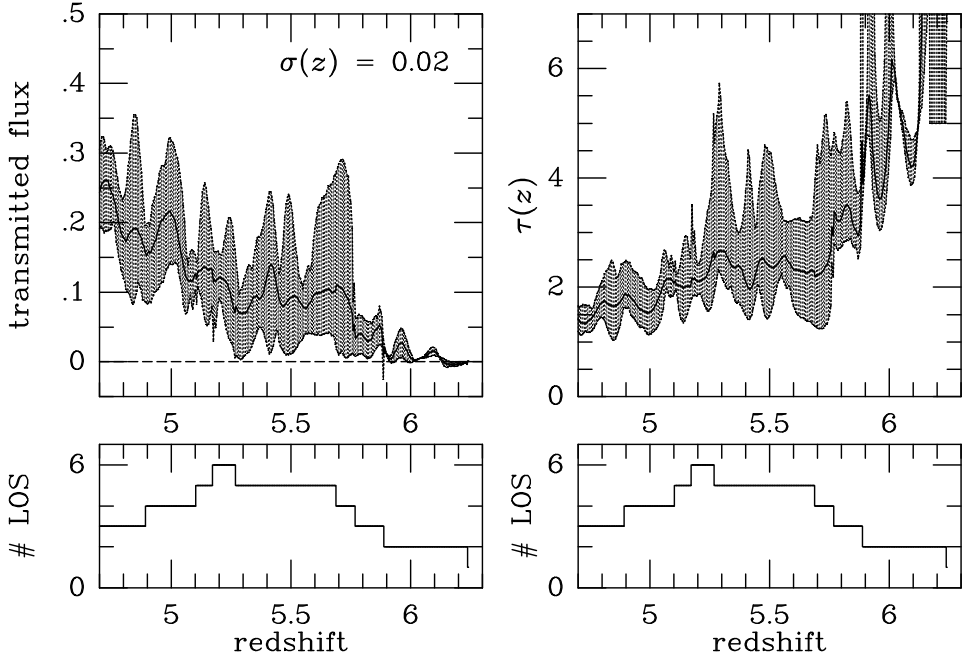,width=5.0in}}
\caption{
The behavior of the transparency of the IGM at high redshifts, shown as the
transmission on the left, and as the optical depth on the right.  The bottom
panels indicate the number of lines of sight used at every redshift.
The solid lines indicate the mean, and the hashed areas the spread among the
different lines of sight.  The data have been smoothed with a Gaussian with
a $\sigma(z) = 0.02$; a wider smoothing or binning, preferred by some authors,
can wash out important features.
Note the change in the slope of $\tau(z)$ around $z \approx 5.8$; this
corresponds to the dramatic change in the ionizing flux and the fraction of
the neutral hydrogen at these redshifts \cite{fan+02}, which
is interpreted as the signature of the end of the reionization era.
There is also a hint that the spread among different lines of sight (i.e.,
the cosmic variance in the IGM absorption properties) increases just before
the reionization redshift is reached.
From~\cite{djo+02,djo+05}.
}
\label{}
\end{figure}

Interestingly, high quality QSO spectra at $z > 6$ can be used to
measure the sizes of the H II regions around these objects, and then to
constrain the neutral hydrogen fractions in the regime 
$x_{H I} \sim 10^{-1} - 10^{0}$~\cite{wl04a,mes+04}
 From a single QSO, Wyithe \& Loeb conclude that $x_{H I} > 10^{-1}$
in its vicinity.  However, this is the $x_{H I}$ range also probed by the
census of the Ly$\alpha$ emitters~\cite{mal+04,ste+05},
where the preliminary results indicate $\langle x_{H I} \rangle < 10^{-1}$
in the general field.  There are both small number statistics and
interpretative uncertainties in both
methods, and clearly at least a few more QSO Str\"omgren sphere measurements
are needed.  If the two approaches continue to give systematically different
results as the data accumulate, that may be
indicative of systematic differences in the large-scale environments, e.g.,
with QSOs being situated in particularly dense regions.

A strong spatial nonuniformity in the final stages of reionization could
be caused by a strong clustering of the reionization sources.  Since there
was not enough time for a substantial large scale structure growth at these
redshifts, the clustering would have to be primordial, e.g., reflecting a
generic biased galaxy formation scenario, where the first luminous sources
form in and around the highest peaks of the primordial density field,
which are a priori strongly clustered almost regardless of any details
of the assumed structure formation models~\cite{kai+84}.  Thus, we expect
that there was a fundamental connection between the early structure formation
and the geometry and topology of the reionization, as reflected in the cosmic
variance of the IGM transmission at these redshifts.

\section{QSOs as Markers of the Primordial Large-Scale Structure}

There is now a substantial and growing evidence for a strong bias among
the field galaxies at redshifts approaching 6; see, e.g.,~\cite{djo+05}
for a review and references.  This is entirely as expected from models of
biased galaxy formation.

Many or all luminous QSOs at these redshifts are also likely situated in
massive host galaxy halos.  This is suggested by the high masses of their
central black holes ($M_\bullet > 10^9 M_\odot$); 
see, e.g.,~\cite{wil+03,die+04,mcl+04};
as well as their highly evolved chemical abundances
\cite{ham+93,ham+99,mat+93,die+03a,die+03b}.
Direct observations of circum-QSO regions at
$z > 4$ show a considerably enhanced density of star forming galaxies in
their vicinity~\cite{djo+98,djo+99a,djo+99b,sti+05}
consistent with the idea that luminous high-$z$ QSOs may mark sites of
future rich clusters of galaxies.

QSOs themselves appear to be strongly clustered at these redshifts;
see~\cite{djo+03a,djo+05} and references therein.  The implied 
clustering lengths are comparable to those of the rich clusters today,
tens of Mpc.
An even more intriguing possibility is that QSO clustering at $z > 4$
extends to physical scales of $\sim 100 - 200 h^{-1}$ comoving Mpc
\cite{djo+98,djo+99b}.
These highly preliminary, uncertain, but
suggestive results must be confirmed by careful analysis of statistical
samples of high-$z$ QSOs.

If QSO clustering at $z \sim 4 - 6$ is real,
it has to be bias-driven.  It would be a nearly unique probe of the
clustering of luminous sources at these redshifts, on physical scales
$\sim 10 - 100$ comoving Mpc, and directly relevant for the interpretation
of the cosmic variance of IGM transmission at the end of the reionization.

Also, if substantial spatial variations in the distribution of the ionized gas
existed at these redshifts, they might have also left an observable imprint
in the CMBR fluctuations, essentially the clustered S-Z effect.  In this
context, it is intriguing to recall the excess power at high angular
frequencies, corresponding to the comoving separations of $\sim 10 - 20$ Mpc
at the CMBR photosphere; see \cite{rea+04} and references therein.

\section{The Palomar-Quest Survey}

As the preceeding discussion indicates, substantially larger samples of
high-$z$ QSOs in the late- and post-recombination redshift regime are
needed in order to provide better observational constraints on the
physical parameters, geometry, and evolution of the IGM in this crucial
era, as well as to quantify better the early, possibly highly biased
large-scale structure.

The Palomar-Quest (PQ) Digital Synoptic Sky Survey 
\cite{pqwebsite, bau+03,rab+03,mah+03,djo+03b,djo+04}
will be a major new provider of high-$z$ QSOs at $z \sim 4 - 6.5$.
The PQ survey is a collaborative
effort between Yale, Caltech, Indiana University and NCSA in the USA,
with collaborations with INAOE in Mexico, and other groups.  The data are 
taken at the 48-inch Samuel Oschin Schmidt telescope at Palomar
Observatory, using a special 112-CCD camera built for this purpose.
The data rate is about 1 TB per month.  Data are reduced and archived
at multiple locations (Caltech, NCSA, Yale).
National Virtual Observatory (NVO)
standards, protocols, and connections are built in from the start, and
will facilitate a very broad data access and analysis.

Approximately 50\% of the time is used for PQ survey drift scans, and
the rest is used largely in the traditional point-and-stare mode by
groups at JPL and Caltech for exploration of the Solar System.
In the drift scan mode, the PQ survey covers strips $\sim 4.6^\circ$
wide at a constant Dec, with an area coverage of $\sim 500\,deg^{2}$
per clear night, in 4 filters.  The pixel scale is 0.878 arcsec.  The
scans are obtained in the range $-25^{\circ} \leq \delta
\leq+25^{\circ}$, giving the total useful survey area of 
$\sim 15,000\,deg^{2}$.  About a third of the survey area overlaps with
that of SDSS, thus providing a valuable cross-check and cross-calibration.
The survey uses two sets of
filters, the traditional $UBRI$, and the SDSS $r'i'z'z'$, with about
an equal time share in each.  Typical limiting magnitudes in a single
pass in good conditions are: $r' \sim 21.5$ mag, $i' \sim 20.5$ mag,
$z' \sim 19.5$ mag, $R \sim 22.0$ mag, and $I \sim 21$ mag.  Coadding
of about 6 passes reaches the depths comparable to SDSS.  As the
survey unfolds, we anticipate that we will cover much of the area
to at least a magnitude deeper, and a subset of the total survey
area deeper yet.

One of the key scientific goals of the PQ survey is the discovery
of a significant number of high-$z$ QSOs, to be used as a statistical
sample for studies of early structure formation and reionization.  The
methodology is identical to that used by SDSS, DPOSS, and other
multi-color high-$z$ QSO surveys to date, using colors as
discriminants for objects classified as point sources in the images.
High-$z$ QSO candidates
are identified in the PQ data using $BRIr'i'z'$ colors, which in
appropriate combinations can be used to find QSOs at $z \sim 4 - 6.5$.
IR photometry is necessary to separate QSO candidates at $z > 5.5$
from late-M and brown dwarfs.  We use coadds of 2MASS images to eliminate
some of the brighter dwarf contaminants, but most candidates
require a deeper IR imaging.  This is currently obtained at the Yale
SMARTS telescopes at CTIO, and at the INAOE/OAGH 2.1-m telescope at
Cananea, Mexico.  The surviving color-selected candidates are then followed
spectroscopically at the Palomar 200-inch telescope, and QSOs at $z > 5$
are selected for deeper and more extensive follow-up studies at the Keck
and other facilities.

The first high-$z$ QSOs have been found, but the survey is still in an early
stage.  We anticipate that we will double the sample of the SDSS high-$z$
QSOs over the next few years.  Studies of the joint SDSS+PQ sample will be
used to provide new observational constraints for structure formation at
$z \sim 4 - 6.5$, and late stages of reionization at $z \sim 6$.

\section{Concluding Comments}

As the reionization becomes complete, the residual H I fraction rapidly drops
over large spatial scales.  QSO absorption spectra are the $only$ currently
known probe of this IGM regime, $\langle x_{H I} \rangle \leq 10^{-2}$.
If this happens at $z \sim 6$, as the data seem to indicate, then 
spectroscopy of QSOs in this redshift regime will remain a unique probe
of the late stages of reionization for some time to come.

Moreover, the data indicate that a considerable cosmic variance is present
in the transmission properties of the IGM at these redshifts, along 
different lines of sight.  If the characteristic physical scales of the
late reionization topology are in the range of tens to hundreds of comoving
Mpc, as the recent theoretical studies suggest \cite{bl04, wl04b, of05},
then statistically significant samples of QSOs will provide the first solid 
observational constraints for theoretical models.

If there is indeed a substantial, bias-driven clustering of the luminous
sources responsible for the reionization (which would then naturally be
spatially clumpy and lead to the observed cosmic variance in the IGM
transmission properties), QSOs may again prove to be useful probes of such
clustering, especially if they are associated with some of the densest
peaks of the density field at the time.  Until galaxy redshift surveys at
$z \sim 6$ start spanning comparable comoving volumes and $>> 100$
comoving Mpc scales -- which may be a while in coming -- QSOs may be
again the only viable probe of such primordial clustering and biasing.

Therefore, we expect that studies of larger high-$z$ samples than currently
available, coming from SDSS, PQ, and other surveys, will continue to
generate ever more significant observational constraints for the models,
leading to a better physical understanding of the final stages of the
reionization era.

\noindent {\bf Acknowledgements.}~~
We are indebted to numerous collaborators, and especially members of the
PQ survey team, for sharing the work and the excitement of a number of
research projects, past, current, and future, on which this paper is
based.  We also thank the staff of Palomar and Keck observatories for
their expert help during numerous observing runs.  This work was
supported in part by the NSF grant AST-0407448, and by the Ajax Foundation.
Processing of the PQ data is supported in part by the NSF grant AST-0326524,
and by resources at NCSA.


\begin{thebibliography}{99}
\frenchspacing

\bibitem{bar+01} Barkana, R., \& Loeb, A., PhysRep {\bf 349}, 128 (2001).

\bibitem{bl04} Barkana, R., \& Loeb, A., ApJ {\bf 609}, 474 (2004).

\bibitem{bau+03} Bauer, A., et al. (The PQ Survey Team), 
        BAAS {\bf 35}, 1262 (2003).

\bibitem{bec+01} Becker, R., et al.~(the SDSS Collaboration), AJ {\bf 122},
        2850 (2001).

\bibitem{bog+06} Bogosavljevic, M., et al., in prep. (2006).

\bibitem{cen+02} Cen, R., \& McDonald, P., ApJ {\bf 570}, 457 (2002).

\bibitem{die+04} Dietrich, M., \& Hamann, F., ApJ {\bf 611}, 761 (2004).

\bibitem{die+03a} Dietrich, M., et al., ApJ {\bf 589}, 722 (2003).
           
\bibitem{die+03b} Dietrich, M., Hamann, F., Appenzeller, I., \& Vestergaard, M.,
        ApJ {\bf 596}, 817 (2003).

\bibitem{djo+98} Djorgovski, S.G., in: Fundamental Parameters in Cosmology,
        eds. Y. Giraud-Heraud et al., Gif sur Yvette: Editions Fronti\`eres,
        p. 313 (1998).

\bibitem{djo+99a} Djorgovski, S.G., Odewahn, S.C., Gal, R.R., Brunner, R., \&
        de Carvalho, R., ASPCS {\bf 191}, 179 (1999).

\bibitem{djo+99b} Djorgovski, S.G., ASPCS {\bf 193}, 397 (1999).

\bibitem{djo+01} Djorgovski, S.G., Castro, S., Stern, D., \& Mahabal, A.
        ApJ {\bf 560}, L5 (2001).

\bibitem{djo+02} Djorgovski, S.G., Bogosavljevic, M., Mahabal, A., Lopes, P.
        \& Castro, S. BAAS {\bf 34}, 1325 (2002).

\bibitem{djo+03a}  Djorgovski, S.G., Stern, D., Mahabal, A., \& Brunner, R. 
        ApJ {\bf 596}, 67 (2003).

\bibitem{djo+03b} Djorgovski, S.G., et al. (The PQ Survey Team), 
        BAAS {\bf 35}, 1315 (2003).

\bibitem{djo+04} Djorgovski, S.G., et al. (The PQ Survey Team), 
        BAAS {\bf 36}, 805 and 1487 (2004).

\bibitem{djo+05} Djorgovski, S.G. in ``Thinking, Observing and Mining the
        Universe'', eds. G. Miele \& G. Longo, Singapore: World Scientific,
        p. 111 (2005) [astro-ph/0409378]

\bibitem{fan+01} Fan, X., et al.~(the SDSS Collaboration), AJ {\bf 122},
        2833 (2001).

\bibitem{fan+02} Fan, X., et al.~(the SDSS Collaboration), AJ {\bf 123},
        1247 (2002).

\bibitem{fan+03} Fan, X., et al.~(the SDSS Collaboration), AJ {\bf 125},
        1649 (2003).

\bibitem{fan+04} Fan, X., et al.~(the SDSS Collaboration), AJ {\bf 128},
        515 (2004).

\bibitem{gun+65} Gunn, J.E., \& Peterson, B.A., ApJ {\bf 142}, 1633 (1965).

\bibitem{ham+93} Hamann, F., \& Ferland, G., ApJ {\bf 418}, 11 (1993).

\bibitem{ham+99} Hamann, F., \& Ferland, G., ARAA {\bf 37}, 487 (1999).

\bibitem{kai+84} Kaiser, N., ApJ {\bf 284}, L9 (1984).

\bibitem{kog+03} Kogut, A., et al.~(the WMAP team), ApJS {\bf 148}, 161 (2003).

\bibitem{loe+01} Loeb, A., \& Barkana, R. ARAA {\bf 39}, 19 (2001).

\bibitem{mah+03} Mahabal, A., et al. (The PQ Survey Team), BAAS {\bf 35},
        1262 (2003).

\bibitem{mal+04} Malhotra, S., \& Rhoads, J., ApJ {\bf 617}, L5 (2004).

\bibitem{mat+93} Matteucci, F., \& Padovani, P., ApJ {\bf 419}, 485 (1993).

\bibitem{mcl+04} McLure, R., \& Dunlop, J., MNRAS {\bf 352}, 1390 (2004).

\bibitem{mes+04} Mesinger, A., \& Haiman, Z. ApJ {\bf 611}, L69 (2004).

\bibitem{mir+03} Miralda-Escud\'e, J., Science {\bf 300}, 1904 (2003).

\bibitem{of05} Oh, S.P. \& Furlanetto, S. ApJ {\bf 620}, L9 (2005).

\bibitem{pqwebsite} The PQ survey website, http://www.astro.caltech.edu/pq/

\bibitem{rab+03} Rabinowitz, D., Baltay, C., et al. (The PQ Survey Team), 
        BAAS {\bf 35}, 1262 (2003).

\bibitem{rea+04} Readhead, A., et al.~(the CBI team), ApJ {\bf 609}, 498 (2004).

\bibitem{ste+05} Stern, D., Yost, S., Eckart, M., Harrison, F., Helfand, D.,
        \& Djorgovski, S.G. ApJ {\bf 619}, 12 (2005).

\bibitem{sti+05} Stiavelli, M., Djorgovski, S.G., Pavlovsky, C., Scarlata, C., Stern, D., 
        Mahabal, A., Thompson, D., Dickinson, M., Panagia, N., \& Meylan, G.,
        ApJ {\bf 622}, L1 (2005).

\bibitem{str+05} Strauss, M., et al. (The SDSS Team),
        presentation at the ``Reionizing the Universe'' conference,
        Groningen, July 2005, available at 
        http://www.astro.rug.nl/~cosmo05/presentations/strauss.pdf

\bibitem{whi+03} White, R., Becker, R., Fan, X., \& Strauss, M., AJ {\bf 126},
        1 (2003).

\bibitem{wil+03} Willott, C., McLure, R., \& Jarvis, M., ApJ {\bf 587}, L15 (2003).

\bibitem{wl04a} Wyithe, S., \& Loeb, A. Nature {\bf 427}, 815 (2004).

\bibitem{wl04b} Wyithe, S., \& Loeb, A. Nature {\bf 432}, 194 (2004).


\end{thebibliography}
\end{document}